\documentclass{aastex631}

\shorttitle{X-Ray Observations of SN 2018cqj}
\shortauthors{Dwarkadas}
\graphicspath{{./}{figures/}}

\begin{document}

\title{X-Ray Observations of Type Ia Supernova 2018cqj}

\correspondingauthor{Vikram V. Dwarkadas}
\email{vikram@astro.uchicago.edu}

\author[0000-0002-4661-7001]{Vikram V. Dwarkadas}
\affiliation{Department of Astronomy and Astrophysics \\
University of Chicago \\
5640 S Ellis Ave., Chicago, IL 60637}



\begin{abstract}

I report on Chandra X-ray observations of SN 2018cqj, a low luminosity Type Ia supernova that showed an H$\alpha$
line in its optical spectrum. No X-ray emission was detected at the location of the SN, with an upper limit to the X-ray luminosity of 2 $\times 10^{39}$ erg s$^{-1}$.  

\end{abstract}



\section{Introduction} \label{sec:intro}

Type Ia supernovae (SNe) have been used to measure the expansion and acceleration of the universe, and thus are important cosmological probes.  Unfortunately the nature of their progenitor systems is still unknown. The progenitor of a Ia is accepted to be a white dwarf that must have accreted mass transferred from a companion in a binary system. The identity of this companion is highly debated \citep{pilar14}, with both single and double degenerate  systems being proposed \citep{maoz14}. In the former case, the companion is another white dwarf \citep{scalzoetal10, nugentetal11}. In the latter it is a main sequence or evolved star \citep{hamuyetal03, dengetal04}.  It is likely that multiple channels exist to form Type Ia SNe \citep{taubenberger17,lm18, rs18, soker19, tn19, ruiter20}.

If a stellar companion is present, mass-loss from the companion can create a H-rich medium around the system, with which the SN shock wave may interact. Alternately, the shock may sweep-up H from the companion. In either case this results in H lines in the optical spectrum of the SN. A subclass of SNe known as Type Ia-CSM SNe ($ \approx 30$ in number) does exhibit narrow hydrogen lines superimposed on a SN Ia-like spectrum \citep{hamuyetal03,dengetal04, silvermanetal13, sharmaetal23}.

The presence of H$\alpha$ in the optical spectrum is suggestive of interaction of the SN shock with a dense surrounding medium. Tracers of this interaction include radio and X-ray emission. Although over 60 SNe have been detected in X-rays \citep{dg12, vvd14, rd17, bocheneketal18}, SN 2012ca, a Type Ia-CSM SN, represents the  only detection \citep{bocheneketal18} of X-ray emission from a Type Ia SN  to date.  

Herein I provide preliminary results from X-ray observations of another Type Ia SN with an H$\alpha$ line in its spectrum.

\section{SN 2018cqj} {\bf SN 2018cqj/ATLAS18qtd} is a low luminosity fast-declining Type Ia in the S0 galaxy IC 550 \citep{prietoetal20}, with a redshift z=0.0165. An H$\alpha$ line was detected in its optical spectrum 193 and 307 days after peak, with FWHM 1200 km s$^{-1}$ after 193 days, and unresolved at 307 d. The line luminosity was $3.8 \pm 0.9 \times 10^{37}$   and $4.6 \pm 1.4 \times 10^{36}$ erg s$^{-1}$ at the two epochs. A lower limit of 6 was estimated for the Balmer decrement. 

Given the presence of the line and the high line luminosities (albeit lower than those of Type Ia-CSM SNe), I proposed for a Chandra observation of this SN. SN 2018cqj was observed by the ACIS-S instrument on the Chandra satellite on Dec 12 2022 (ObsID 26613, PI Dwarkadas), 1643 days after discovery (1616 in SN rest frame), for a total exposure time of 44.88 ks. The data were downloaded and analysed using CIAO 4.15 and CalDB 4.10.2. 

A 1$^{\prime\prime}$ source region was used. For the background I used an annulus of inner radius 1$^{\prime\prime}$ and outer radius 5$^{\prime\prime}$. The source region includes 1 count between 0.5-8 keV, with 8 counts in the background.  I use the Bayesian method of \citet{kbn91} to determine the maximum number of counts, deriving a maximum of 7.6 counts at a 99.7\% (3$\sigma$) confidence level, and a count rate 1.7 $\times 10^{-4}$ counts s$^{-1}$. Using a Galactic column density towards the source of 3.41 $\times 10^{20}$ cm$^{-2}$ \citep{dl90}, and Chandra PIMMS, I determine the flux for a power-law model with index 2, appropriate for inverse-Compton emission, to be 2.67 $\times 10^{-15}$ erg s$^{-1}$ cm$^{-2}$. The {\sc srcflux} command in Sherpa for the same model returns a maximum flux of 2.9 $\times 10^{-15}$ erg s$^{-1}$ cm$^{-2}$ (Table 1).

If the density around the SN is high, as suggested by the H$\alpha$ line, then the emission may be thermal, due to thermal bremsstrahlung combined with line emission. I also  calculate the flux using the Plasma/APEC model, for a range of temperatures, using a Galactic column density. This returns a flux $< 2.5 \times 10^{-15}$ erg s$^{-1}$ cm$^{-2}$ (Table 1). 

\begin{deluxetable*}{lcCC}
\tablenum{1}
\tablecaption{Flux and Luminosity Calculations for SN 2018cqj\label{tab:18cqj}}
\tablewidth{0pt}
\tablehead{
\colhead{Model} & \colhead{temperature}  & \colhead{Flux} &
 \colhead{Luminopsity} \\
\colhead{} & \colhead{keV} &\colhead{erg s$^{-1}$ cm$^{-2}$} &
 \colhead{(erg s$^{-1}$)}
}
\decimalcolnumbers
\startdata
Power-Law ($\Gamma$=2) &  & < 2.67 \times 10^{-15} & < 1.8 \times 10^{39}  \\
(Using {\sc srcflux}) &   & <  2.90 \times 10^{-15} & < 1.9 \times 10^{39}   \\
Plasma/APEC &  1.53 & < 2.17 \times 10^{-15} & < 1.4 \times 10^{39}   \\
Plasma/APEC &  3.1 & < 2.37 \times 10^{-15} & < 1.6 \times 10^{39}   \\
Plasma/APEC &  4.3 & < 2.52 \times 10^{-15} & < 1.7 \times 10^{39}   \\
\enddata
\tablecomments{The unabsorbed flux and luminosity of SN 2018cqj, assuming  a distance of 74.3 Mpc.
3$\sigma$ results are quoted.}
\end{deluxetable*}

\section{Conclusions} SN 2018cqj was not detected in an $\approx$ 45 ks observation with Chandra ACIS-S. The derived upper limits suggest an intrinsic X-ray luminosity less than  $ 2 \times10^{39}$ erg s$^{-1}$ for a distance of 74.3 Mpc. This implies that if the SN was expanding in a region of high density at the time the H$\alpha$ line was seen, the density had decreased by the time of the Chandra observation.

The optical properties of SN 2018cqj were similar to those of SN 2018fhw/ASASSN-18tb, another low-luminosity Type Ia with an H$\alpha$ line in the spectrum. A Chandra observation of SN 2018fhw in July 2021 detected no X-ray emission \citep{vvd23}, with an upper limit to the luminosity of 3 $\times 10^{39}$ erg s$^{-1}$.

\begin{acknowledgments}
Support for this work was provided by NASA through Chandra
Award GO3-24041X  issued by the Chandra X-ray Center (CXC), operated by SAO
for and on behalf of NASA under contract NAS8-03060; and by NSF
grant 1911061. Scientific results reported  are based
 on observations made by the Chandra X-ray Observatory (ObsID 26613). This
research has made use of CXC software  CIAO and Sherpa, as well as Chandra PIMMS and Colden packages. 
\end{acknowledgments}

%

\vspace{5mm}
\facilities{Chandra (ACIS-S)}


\software{CIAO \citep{ciaocite}, Sherpa \citep{sherpacite}, PIMMS (https://cxc.harvard.edu/toolkit/pimms.jsp), Colden (https://cxc.harvard.edu/toolkit/colden.jsp)}



\bibliography{halpha}{}

\begin{thebibliography}{}
\expandafter\ifx\csname natexlab\endcsname\relax\def\natexlab#1{#1}\fi
\providecommand{\url}[1]{\href{#1}{#1}}
\providecommand{\dodoi}[1]{doi:~\href{http://doi.org/#1}{\nolinkurl{#1}}}
\providecommand{\doeprint}[1]{\href{http://ascl.net/#1}{\nolinkurl{http://ascl.net/#1}}}
\providecommand{\doarXiv}[1]{\href{https://arxiv.org/abs/#1}{\nolinkurl{https://arxiv.org/abs/#1}}}

\bibitem[{{Bochenek} {et~al.}(2018){Bochenek}, {Dwarkadas}, {Silverman}, {Fox},
  {Chevalier}, {Smith}, \& {Filippenko}}]{bocheneketal18}
{Bochenek}, C.~D., {Dwarkadas}, V.~V., {Silverman}, J.~M., {et~al.} 2018,
  \mnras, 473, 336, \dodoi{10.1093/mnras/stx2029}

\bibitem[{{Deng} {et~al.}(2004){Deng}, {Kawabata}, {Ohyama}, {Nomoto},
  {Mazzali}, {Wang}, {Jeffery}, {Iye}, {Tomita}, \& {Yoshii}}]{dengetal04}
{Deng}, J., {Kawabata}, K.~S., {Ohyama}, Y., {et~al.} 2004, \apjl, 605, L37,
  \dodoi{10.1086/420698}

\bibitem[{{Dickey} \& {Lockman}(1990)}]{dl90}
{Dickey}, J.~M., \& {Lockman}, F.~J. 1990, \araa, 28, 215,
  \dodoi{10.1146/annurev.aa.28.090190.001243}

\bibitem[{{Dwarkadas}(2014)}]{vvd14}
{Dwarkadas}, V.~V. 2014, \mnras, 440, 1917, \dodoi{10.1093/mnras/stu347}

\bibitem[{{Dwarkadas}(2023)}]{vvd23}
---. 2023, \mnras, 520, 1362, \dodoi{10.1093/mnras/stac3384}

\bibitem[{{Dwarkadas} \& {Gruszko}(2012)}]{dg12}
{Dwarkadas}, V.~V., \& {Gruszko}, J. 2012, \mnras, 419, 1515,
  \dodoi{10.1111/j.1365-2966.2011.19808.x}

\bibitem[{{Freeman} {et~al.}(2001){Freeman}, {Doe}, \&
  {Siemiginowska}}]{sherpacite}
{Freeman}, P., {Doe}, S., \& {Siemiginowska}, A. 2001, in SPIE Conference
  Series, Vol. 4477, Astronomical Data Analysis, ed. J.-L. {Starck} \& F.~D.
  {Murtagh}, 76--87, \dodoi{10.1117/12.447161}

\bibitem[{{Fruscione} {et~al.}(2006){Fruscione}, {McDowell}, {Allen},
  {Brickhouse}, {Burke}, {Davis}, {Durham}, {Elvis}, {Galle}, {Harris},
  {Huenemoerder}, {Houck}, {Ishibashi}, {Karovska}, {Nicastro}, {Noble},
  {Nowak}, {Primini}, {Siemiginowska}, {Smith}, \& {Wise}}]{ciaocite}
{Fruscione}, A., {McDowell}, J.~C., {Allen}, G.~E., {et~al.} 2006, in SPIE
  Conference Series, Vol. 6270, SPIE Conference Series, ed. D.~R. {Silva} \&
  R.~E. {Doxsey}, 62701V, \dodoi{10.1117/12.671760}

\bibitem[{{Hamuy} {et~al.}(2003){Hamuy}, {Phillips}, {Suntzeff}, {Maza},
  {Gonz{\'a}lez}, {Roth}, {Krisciunas}, {Morrell}, {Green}, {Persson}, \&
  {McCarthy}}]{hamuyetal03}
{Hamuy}, M., {Phillips}, M.~M., {Suntzeff}, N.~B., {et~al.} 2003, \nat, 424,
  651, \dodoi{10.1038/nature01854}

\bibitem[{{Kraft} {et~al.}(1991){Kraft}, {Burrows}, \& {Nousek}}]{kbn91}
{Kraft}, R.~P., {Burrows}, D.~N., \& {Nousek}, J.~A. 1991, \apj, 374, 344,
  \dodoi{10.1086/170124}

\bibitem[{{Livio} \& {Mazzali}(2018)}]{lm18}
{Livio}, M., \& {Mazzali}, P. 2018, \physrep, 736, 1,
  \dodoi{10.1016/j.physrep.2018.02.002}

\bibitem[{{Maoz} {et~al.}(2014){Maoz}, {Mannucci}, \& {Nelemans}}]{maoz14}
{Maoz}, D., {Mannucci}, F., \& {Nelemans}, G. 2014, \araa, 52, 107,
  \dodoi{10.1146/annurev-astro-082812-141031}

\bibitem[{{Nugent} {et~al.}(2011){Nugent}, {Sullivan}, {Cenko}, {Thomas},
  {Kasen}, {Howell}, {Bersier}, {Bloom}, {Kulkarni}, {Kandrashoff},
  {Filippenko}, {Silverman}, {Marcy}, {Howard}, {Isaacson}, {Maguire},
  {Suzuki}, {Tarlton}, {Pan}, {Bildsten}, {Fulton}, {Parrent}, {Sand},
  {Podsiadlowski}, {Bianco}, {Dilday}, {Graham}, {Lyman}, {James}, {Kasliwal},
  {Law}, {Quimby}, {Hook}, {Walker}, {Mazzali}, {Pian}, {Ofek}, {Gal-Yam}, \&
  {Poznanski}}]{nugentetal11}
{Nugent}, P.~E., {Sullivan}, M., {Cenko}, S.~B., {et~al.} 2011, \nat, 480, 344,
  \dodoi{10.1038/nature10644}

\bibitem[{{Prieto} {et~al.}(2020){Prieto}, {Chen}, {Dong}, {Bose}, {Gal-Yam},
  {Holoien}, {Kollmeier}, {Phillips}, \& {Shappee}}]{prietoetal20}
{Prieto}, J.~L., {Chen}, P., {Dong}, S., {et~al.} 2020, \apj, 889, 100,
  \dodoi{10.3847/1538-4357/ab6323}

\bibitem[{{R{\"o}pke} \& {Sim}(2018)}]{rs18}
{R{\"o}pke}, F.~K., \& {Sim}, S.~A. 2018, \ssr, 214, 72,
  \dodoi{10.1007/s11214-018-0503-8}

\bibitem[{{Ross} \& {Dwarkadas}(2017)}]{rd17}
{Ross}, M., \& {Dwarkadas}, V.~V. 2017, \aj, 153, 246,
  \dodoi{10.3847/1538-3881/aa6d50}

\bibitem[{{Ruiter}(2020)}]{ruiter20}
{Ruiter}, A.~J. 2020, IAU Symposium, 357, 1, \dodoi{10.1017/S1743921320000587}

\bibitem[{{Ruiz-Lapuente}(2014)}]{pilar14}
{Ruiz-Lapuente}, P. 2014, \nar, 62, 15, \dodoi{10.1016/j.newar.2014.08.002}

\bibitem[{{Scalzo} {et~al.}(2010){Scalzo}, {Aldering}, {Antilogus}, {Aragon},
  {Bailey}, {Baltay}, {Bongard}, {Buton}, {Childress}, {Chotard}, {Copin},
  {Fakhouri}, {Gal-Yam}, {Gangler}, {Hoyer}, {Kasliwal}, {Loken}, {Nugent},
  {Pain}, {P{\'e}contal}, {Pereira}, {Perlmutter}, {Rabinowitz}, {Rau},
  {Rigaudier}, {Runge}, {Smadja}, {Tao}, {Thomas}, {Weaver}, \&
  {Wu}}]{scalzoetal10}
{Scalzo}, R.~A., {Aldering}, G., {Antilogus}, P., {et~al.} 2010, \apj, 713,
  1073, \dodoi{10.1088/0004-637X/713/2/1073}

\bibitem[{{Sharma} {et~al.}(2023){Sharma}, {Sollerman}, {Fremling}, {Kulkarni},
  {De}, {Irani}, {Schulze}, {Strotjohann}, {Gal-Yam}, {Maguire}, {Perley},
  {Bellm}, {Kool}, {Brink}, {Bruch}, {Deckers}, {Dekany}, {Dugas},
  {Filippenko}, {Goldwasser}, {Graham}, {Graham}, {Groom}, {Hankins},
  {Jencson}, {Johansson}, {Karambelkar}, {Kasliwal}, {Masci}, {Medford},
  {Neill}, {Nir}, {Riddle}, {Rigault}, {Schweyer}, {Terwel}, {Yan}, {Yang}, \&
  {Yao}}]{sharmaetal23}
{Sharma}, Y., {Sollerman}, J., {Fremling}, C., {et~al.} 2023, \apj, 948, 52,
  \dodoi{10.3847/1538-4357/acbc16}

\bibitem[{{Silverman} {et~al.}(2013){Silverman}, {Nugent}, {Gal-Yam},
  {Sullivan}, {Howell}, {Filippenko}, {Arcavi}, {Ben-Ami}, {Bloom}, {Cenko},
  {Cao}, {Chornock}, {Clubb}, {Coil}, {Foley}, {Graham}, {Griffith}, {Horesh},
  {Kasliwal}, {Kulkarni}, {Leonard}, {Li}, {Matheson}, {Miller}, {Modjaz},
  {Ofek}, {Pan}, {Perley}, {Poznanski}, {Quimby}, {Steele}, {Sternberg}, {Xu},
  \& {Yaron}}]{silvermanetal13}
{Silverman}, J.~M., {Nugent}, P.~E., {Gal-Yam}, A., {et~al.} 2013, \apjs, 207,
  3, \dodoi{10.1088/0067-0049/207/1/3}

\bibitem[{{Soker}(2019)}]{soker19}
{Soker}, N. 2019, \nar, 87, 101535, \dodoi{10.1016/j.newar.2020.101535}

\bibitem[{{Tanikawa} {et~al.}(2019){Tanikawa}, {Nomoto}, {Nakasato}, \&
  {Maeda}}]{tn19}
{Tanikawa}, A., {Nomoto}, K., {Nakasato}, N., \& {Maeda}, K. 2019, \apj, 885,
  103, \dodoi{10.3847/1538-4357/ab46b6}

\bibitem[{{Taubenberger}(2017)}]{taubenberger17}
{Taubenberger}, S. 2017, in Handbook of Supernovae, ed. A.~W. {Alsabti} \&
  P.~{Murdin}, 317, \dodoi{10.1007/978-3-319-21846-5\_37}

\end{thebibliography}
\bibliographystyle{aasjournal}



\end{document}